\documentclass[conference,letterpaper,comsoc]{IEEEtran}

\usepackage[utf8]{inputenc} 
\usepackage[T1]{fontenc}
\usepackage{url}
\usepackage{ifthen}
\usepackage{cite}
\usepackage[cmex10]{amsmath} 

\usepackage{amsmath,amsfonts,comment,amssymb}
\usepackage[pdftex]{graphicx}
\newtheorem{theorem}{Theorem}

\newtheorem{definition}[theorem]{Definition}
\newtheorem{algorithm}[theorem]{Algorithm}

\newcommand{\type}[1]{\emph{#1}}

\begin{document}

\title{Rates of Adaptive Group Testing \\ in the Linear Regime}

\author{%
  \IEEEauthorblockN{Matthew Aldridge}
  \IEEEauthorblockA{School of Mathematics\\
                 University of Leeds\\
                 Email: m.aldridge@leeds.ac.uk}
}

\maketitle

\begin{abstract}
We consider adaptive group testing in the linear regime, where the number of defective items scales linearly with the number of items. We analyse an algorithm based on generalized binary splitting. Provided fewer than half the items are defective, we achieve rates of over 0.9 bits per test for combinatorial zero-error testing, and over 0.95 bits per test for probabilistic small-error testing.
\end{abstract}


\section{Introduction}

Group testing is this problem: Given a collection of items some of which are defective, how many pooled tests are required to recover the defective set? A pooled test is performed on some subset of the items: the test is negative if all items in the test are nondefective, and is positive if at least one item in the test is defective.

In Dorfman's original work \cite{dorfman}, the application was to test men enlisting into the U.S.~army for syphilis using a blood test. Dorfman noted that testing pools of mixed blood samples could use fewer tests than testing each blood sample individually. The test result from such a pool should be negative if every blood sample in the pool is free of the disease, while the test result should be positive if at least one of the blood samples is contaminated. 

Different group testing models are discussed in the recent surey \cite{survey}. The most important distinction between is between:
\begin{itemize}
  \item \type{Adaptive testing}, where the items placed in a test can depend on the results of previous tests.
  \item \type{Nonadaptive testing}, where all the tests are designed in advance.
\end{itemize}
This paper concerns adaptive testing, and will examine some cases where adaptive group testing provides large improvements over the nonadaptive case.

Another consideration is how many defective items there are. In this paper, we consider the \emph{linear regime}, where the number of defective items $k$ is a constant proportion $p \in (0,1)$ of the $n$ items. A lot of group testing work has concerned the \emph{very sparse regime} where $k$ is constant as $n \to \infty$ \cite{dyachkov-rykov,malyutov,AS} or the \emph{sparse regime} $k = \Theta(n^\alpha)$  as $n \to \infty$ for some $\alpha < 1$ \cite{BJA,ABJ,SC}. However, we argue that the linear regime is more appropriate for many applications. For example, in Dorfman's original set-up, we might expect each person joining the army to have a similar prior probability $p$ of having the disease, and that this probability should remain roughly constant as more people join, rather than tending towards $0$; thus one expects $k \approx pn$ to grow linearly with $n$.

For group testing in the linear regime, two cases have received most consideration in the literature:
\begin{itemize}
\item \type{Combinatorial zero-error testing:} The defective set is any subset of $\{1,2,\dots,n\}$ with given size $k$, and one wishes to find the defective set with certainty, whichever such set it is. One assumes that $k/n$ tends to a constant $p \in (0,1)$ as $n \to\infty$. \cite{HHW,RC,huang}
\item \type{Probabilistic small-error testing:} We assume each item is defective with probability $p$, independent of all other items, where $p \in (0,1)$ stays fixed as $n \to \infty$. We want to find the defective set with arbitrarily small error probability (averaged over the random defective set). \cite{ungar,mezard,AJM,aldridge}
\end{itemize}




For group testing in the linear regime, it is easy to see that the optimal scaling is the number of tests $T$ scaling linearly with $n$. A simple counting bound (see, for example, \cite{BJA}) shows that we require $T \geq H(p)n$ for large enough $n$, where $H(p)$ is the binary entropy. Meanwhile, testing each item individually requires $T = n$ tests, and succeeds with certainty. (In the combinatorial case, $T = n-1$ suffices, as the status of the final item can be inferred from whether $k$ or $k-1$ defective items have been already discovered from individual tests.) 
The goal of this paper is to analyse algorithms that require a number of tests very close to the lower bound $H(p)n$.

In the sparse regime $k = \Theta(n^\alpha)$ for $\alpha \in [0,1)$, it is known that adaptive testing achieves the counting bound, for both small-error and zero-error criteria, using the generalised binary splitting algorithm of Hwang \cite{hwang,du-hwang,BJA}. 

For nonadaptive testing in the linear regime, it is well known that individual testing is optimal for all $p \in (0,1)$ in the combinatorial zero-error case \cite{dyachkov-rykov,chen-hwang,huang,du-hwang}, and it was recently shown that this is also the case for probabilistic small-error testing too \cite{aldridge}. Thus, for small $p$, the benefit provided by the adaptive algorithms of this paper will be considerable.

Adaptive group testing in the linear regime has received some attention in the literature. The main point of study has been the question of when individual testing is optimal or not. In the combinatorial zero-error case, Riccio and Colburn  \cite{RC} showed that individual testing cannot be improved on for
$p > 1 - \log_3 2 \approx 0.369$, and it is conjectured that this holds for $p > 1/3$ \cite{HHW}.
In the probabilistic case, if one considers the \emph{average} number of tests required, Ungar \cite{ungar} showed that individual testing cannot be improved on for
  \[ p > p^* := \frac{3 - \sqrt{5}}{2} \approx 0.382 . \]
In the linear regime, we are not aware of any work that has aimed to get a number of tests close to optimal over the whole range of $p$, as we do here. 
(Zaman and Pippenger \cite{zaman} do consider this in the limit as $p\to0$.)
Another novelty of ours is that we analyse small-error behaviour, not just average-case behaviour, which allows a direct comparison to nonadaptive results.

\begin{figure*} 
\begin{center}
\includegraphics[width=0.47\textwidth]{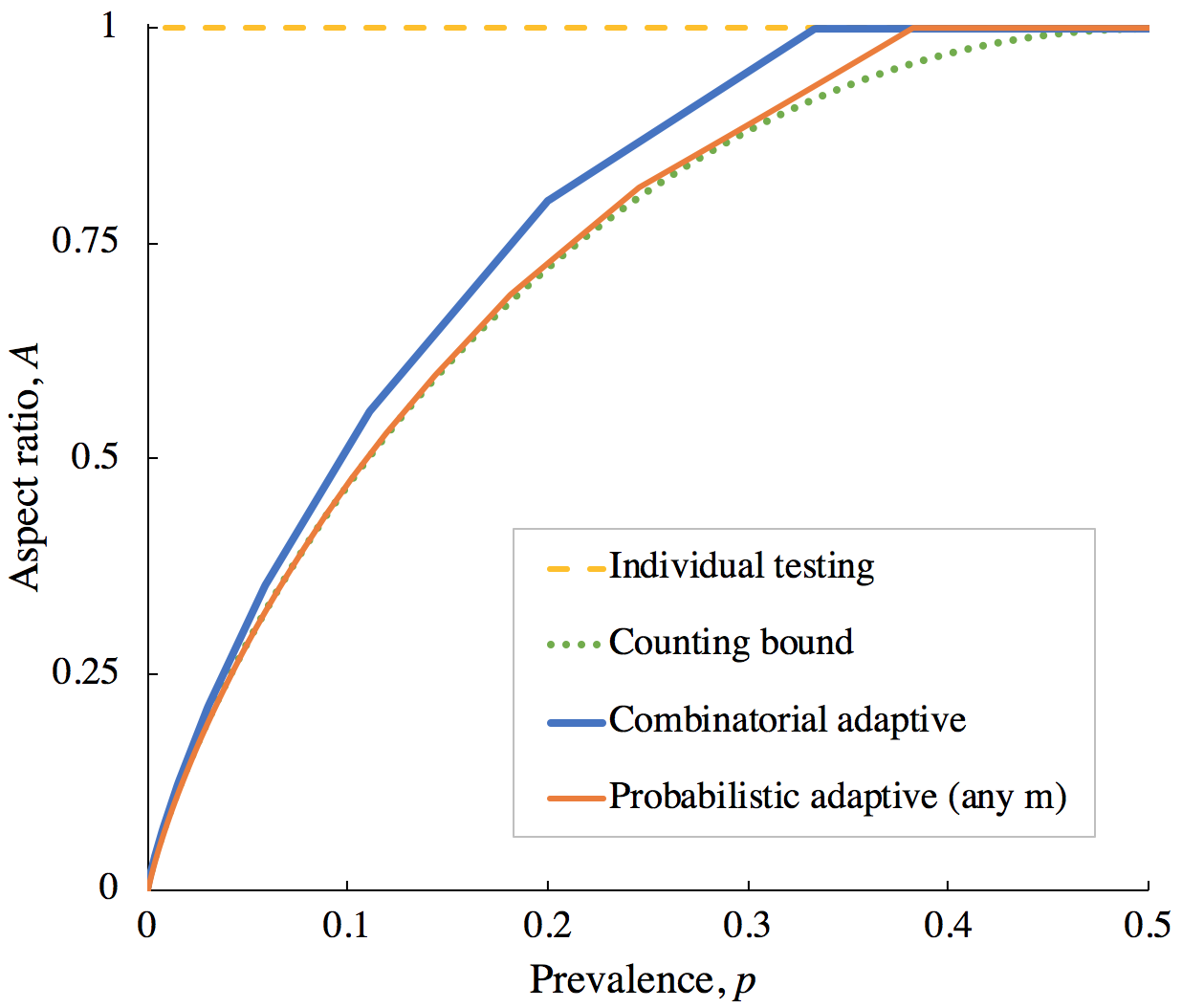}\qquad 
\includegraphics[width=0.47\textwidth]{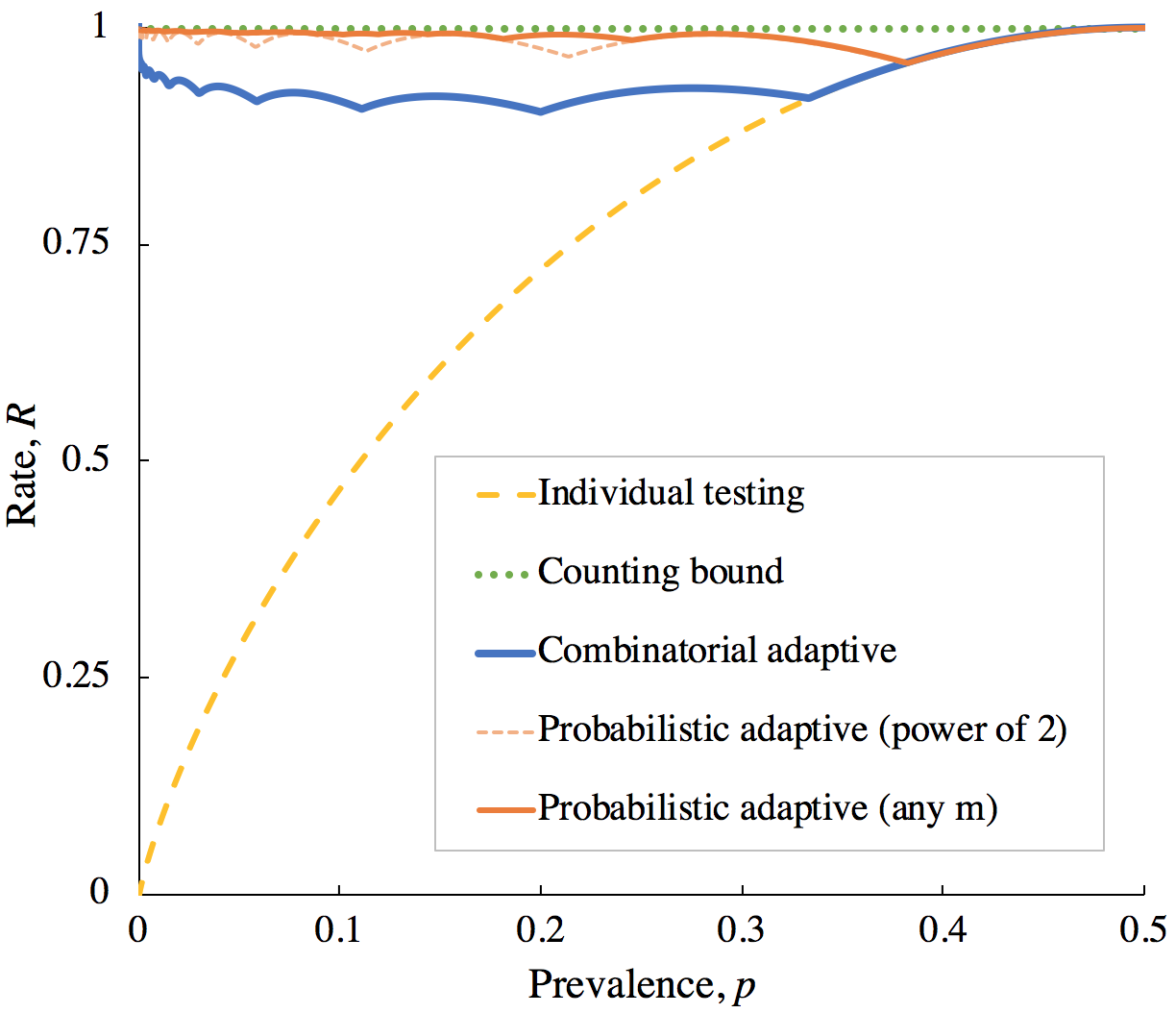}
\end{center}
\vspace{-7px}
\caption{Achievable aspect ratios \emph{(left)} and rates \emph{(right)} for Algorithm \ref{mainalg}, according to Theorems \ref{mainthm1} and \ref{mainthm2}.} \label{mainfig}
\end{figure*}

The goal of this paper is to achieve performance that is close to optimal for adaptive testing under both the zero-error and small-error criteria. We do this using an algorithm similar to that of Hwang \cite{hwang} (see Algorithm \ref{mainalg}), and examining both its worst-case and average-case behaviour. Recall that the counting bound tells us we require at least $T \geq H(p)n$ tests. Our main results are the following, which show very close to optimal performance:
\begin{itemize}
  \item In the zero-error case, we give an algorithm that uses $T < 1.11 H(p)n$ tests for all $p \leq \frac12$. (Theorem \ref{mainthm1})
  \item In the small-error case, we give an algorithm that uses $T < 1.05 H(p)n$ tests for all $p \leq \frac12$. (Theorem \ref{mainthm2})
\end{itemize}

\section{Definitions and main results}

We propose two figures of merit for assessing group testing in the linear regime.

First we have the \emph{aspect ratio} $A = T/n$ (as considered by, for example, \cite{AJM}). We want the aspect ratio to be as small as possible. Individual testing achieves $A = 1$, while the counting bound tells us that we must have $A \geq H(p)$.

Second, we have the \emph{rate} $H(\mathcal K)/T$, where $H(\mathcal K)$ is the entropy of the defective set (as considered by \cite{BJA,ABJ,KJP} and many others). Since $H(\mathcal K)$ is the number of bits required to define the defective set, we can think of the rate as the average number of bits of information learned per test. For combinatorial testing in the linear regime we have
  \[ H(\mathcal K) = \log_2 \binom nk \sim nH\left(\frac kn\right) \sim nH(p) \]
asymptotically, while for probabilistic testing $H(\mathcal K) = nH(p)$ exactly. Hence we can define the rate to be
  \[ R = \frac{n H(p)}{T} = \frac{H(p)}{A} . \]
We want the rate to be as big as possible. Individual testing achieves $R = H(p)$, while the counting bound tells us that we must have $R \leq 1$.

As a rule of thumb, we recommend the aspect ratio for measuring how much better an algorithm is than individual testing, and recommend the rate for measuring how close an algorithm is to the counting bound or comparing with results from the sparse regime.

\begin{definition}
We say that an aspect ratio $A$ is \emph{zero-error achievable} if there is an algorithm with aspect ratio $T/n \geq A$ and error probability $0$ for $n$ sufficiently large. We say that $A$ is \emph{average-case} achievable if there is an algorithm with average-case aspect ratio $\bar T/n \geq A$ and error probability $0$ for $n$ sufficiently large. We say that $A$ is \emph{small-error achievable} if, for any $\delta > 0$, there exists an algorithm with aspect ratio $T/n \geq A$ and average error probability less than $\delta$ for $n$ sufficiently large.
\end{definition}

The equivalent definitions hold for achievable rates, \emph{mutatis mutandis}.

We now state our two main results. We write $\lfloor x \rfloor$ for the greatest integer less than or equal to $x$, and $\lfloor x \rfloor_2 = 2^{\lfloor \log_2 x \rfloor}$ for the greatest power of $2$ less than or equal to $x$; so $\lfloor 5.7 \rfloor = 5$ and $\lfloor 5.7 \rfloor_2 = 4$. We write $q = 1-p$.

\begin{theorem} \label{mainthm1}
Consider nonadaptive group testing in the linear regime. Using Algorithm \ref{mainalg}, all aspect ratios up to
  \[ A =  \frac1m + \left(1 + \log_2 m - \frac1m\right) p , \]
and rates up to $H(p)/A$ are zero-error achievable, where
\[ m= \left\lfloor \frac1p - 1 \right\rfloor_2 . \]
\end{theorem}

\begin{theorem} \label{mainthm2}
Consider nonadaptive group testing in the linear regime. Using Algorithm \ref{mainalg}, all aspect ratios up to
  \[ A = \frac{q^{m} + (1 - q^{m-2b})(a+1)  + (q^{m-2b} - q^m)(a+2)}{{mq^m + \frac{1}{p} \big(1 + mq^{m+1} - (m+1)q^m\big)}} , \]
and rates up to $H(p)/A$ are small-error achievable, where
\[ m = \left\lceil -\frac{\log(2-p)}{\log(1-p)} \right\rceil , \quad a = \lfloor \log_2 m \rfloor, \quad b = m - \lfloor m \rfloor_2 . \]
\end{theorem}

These aspect ratios and rates are illustrated in Fig.~\ref{mainfig}. Note that, for zero-error, we have $R > 0.9$ for all $p \leq 1/2$, and for small-error, $R > 0.95$ for all $p \leq 1/2$. The `bumpy' behaviour occurs from when the optimal value of $m$ switches to the next integer or power of $2$.

\section{Algorithm}

Our algorithm is based on the idea of binary splitting. Binary splitting was first introduced for group testing by Sobel and Groll \cite{sobel}, and our algorithms here are inspired by Hwang's generalized binary splitting \cite{hwang}.

Binary splitting is particularly simple when the size of the set is known to be a power of $2$.

\begin{algorithm} \label{pow2}
Let $\mathcal B$ be a set of items known to contain at least one defective item. Suppose $|\mathcal B| = m$ where $m$ is a power of $2$.
\begin{enumerate}
    \item If $|\mathcal B| = 1$, then that item is defective. Stop.
    \item Otherwise, let $\mathcal C$ consist of the first $|\mathcal B|/2$ items of $\mathcal B$. Test $\mathcal C$.
    \begin{enumerate}
        \item If the test is positive: Set $\mathcal B := \mathcal C$, and return to step 1.
        \item If the test is negative: All items in $\mathcal C$ are nondefective. Set $\mathcal B := \mathcal B \setminus \mathcal C$, and return to step 1.
    \end{enumerate}
\end{enumerate}
\end{algorithm}

\smallskip

Binary splitting where $m$ is a power of $2$ will suffice to prove the most important claims of this paper, of rates above $0.9$ and $0.95$ for zero- and small-error respectively. However, in the small-error case, for some $p < 1/4$ it will be possible to slightly improve the rate by allowing $m$ to be any integer. We postpone discussion of this until Section \ref{sec:gen}.

We now explain our main algorithm.

\begin{algorithm} \label{mainalg}
Let $\mathcal A = \{1,2,\dots,n\}$ be the set of items. We fix an integer parameter $m$.
\begin{enumerate}
  \item If $|\mathcal A| < m$, test the items individually, then halt.
  \item Otherwise, remove the first $m$ items from $\mathcal A$, and call these items $\mathcal B$. Test $\mathcal B$. 
  \begin{enumerate}
      \item If the test is negative: All items in $\mathcal B$ are nondefective. Return to step 1.
      \item If the test is positive: Perform binary splitting on $\mathcal B$  (using Algorithm \ref{pow2} is $m$ is a power of $2$ and Algorithm \ref{gen} otherwise). This will discover $1$ defective item and between $0$ and $m-1$ nondefective items. Return the remaining items whose statuses are not discovered to $\mathcal A$. Return to step 1.
  \end{enumerate}
\end{enumerate}
\end{algorithm}

\smallskip

Since we will choose $m$ independently of $n$ and consider asymptotics as $n \to \infty$, the small number of individual tests incurred at step 1 (which will happen at the end of the algorithm) will be negligible for our calculations here, so we will ignore them in our analysis.

We note that, in the special case $m = 1$, this algorithm is equivalent to individual testing; while in the special case $m = 2$, we recover an algorithm studied by Fischer, Klasner and Wegenera \cite{FKW}. We discuss connections with the work of Zaman and Pippenger \cite{zaman} in Section \ref{sec:gen}.

\section{Worst-case analysis and zero-error rate}

We will use a worst-case analysis of our algorithm to find a zero-error achievable aspect ratio.

\begin{IEEEproof}[Proof of Theorem \ref{mainthm1}]
We perform Algorithm \ref{mainalg} with $m$ a power of $2$ to be fixed later.

In each pass through step 2 of Algorithm \ref{mainalg}, one of two things can happen:
\begin{enumerate}
  \item[a)] The set contains no defectives, in which case we discover $m$ nondefectives with $1$ test.
  \item[b)] The set contains at least one defective, in which case we discover $1$ defective and between $0$ and $m-1$ nondefectives with $1 + \log_2 m$ tests. 
\end{enumerate}

For the purposes of worst-case analysis, we assume that in the second case, we never get lucky, and only ever find the $1$ defective with $0$ bonus nondefectives.
Thus in our $T$ tests we must discover all $n-k$ nondefectives from case 1 and all $k$ defectives from case 2. This gives a worst-case number of tests as
  \[ T =  \frac{1}{m} (n - k) + (1 + \log_2 m)k . \]
This has an aspect ratio of
  \[ A = \frac{1}{m}(1-p) + (1 + \log_2 m)p = \frac1m + \left(1 + \log_2 m - \frac1m\right) p . \]
  
Choosing $m$ as in the statement of the theorem gives the result, and this is easily checked to be the optimal choice of $m$.
\end{IEEEproof}

When $m = 1$, we have individual testing with
  \[ T = (n-k) + k = n . \]
When $m = 2$, we have 
  \[ T = \frac{1}{2} (n - k) + 2k = \frac12 n + \frac 32 k = \left( \frac12 + \frac32 p \right) n, \]
recovering a result of \cite{FKW}. The $m=2$ case beats individual testing when $p < 1/3$, recovering a result of Hu, Hwang and Wang \cite{HHW}, also noted in \cite{FKW}.  

\section{Average-case analysis and small-error rate}

To get a small-error achievability result, we start with an average-case analysis, and later twin this with a concentration of measure argument.

\subsection{Powers of $2$ algorithm: average-case analysis}

We begin with average-case analysis of the simpler case when $m$ is a power of $2$.


Again, we look at the outcomes for a pass through step 2.
\begin{enumerate}
  \item With probability $q^m$, all items in the test are negative, and we discover their nondefective statuses with $1$ test.
  \item With probability $1 - q^m$ there is at least one defective in the test. Let $j$ be the first-numbered defective in the set. We discover defective status of item $j$ and the nondefective statuses of items $1,2,\dots,j-1$ in $1 + \log_2 m$ tests.
\end{enumerate}

The expected number of tests in one pass through step 2 is
  \[ F = q^m\cdot 1 + (1 - q^m)(1 + \log_2 m) =  1 +(1 - q^m)\log_2 m . \]
The expected number of items whose status we discover is
  \begin{align*}
    G &= mq^m + \sum_{j=1}^m jp q^{j-1}  \notag \\
      &= mq^m + \frac{1}{p} \big(1 + mq^{m+1} - (m+1)q^m\big)  .
  \end{align*}
(The sum here has an explicit form since $\sum_j  jq^{j-1} = \frac{\mathrm{d}}{\mathrm{d}j} \sum_j q^j$.)
  
Since the average aspect ratio $A$ is the ratio of the average number of tests to the number of items, it seems plausible that $A = F/G$. To prove this rigorously, note that the number of tests the algorithm takes on average is, by considering one pass through step 2,
  \begin{align} An = \mathbb E\, T  &= \mathbb E \,\#\,\text{tests performed on one pass} \notag \\
  &\qquad {}+ \mathbb E\,\#\,\text{tests to deal with all remaining items} \notag \\
  &= F + A\,\mathbb E\,\#\,\text{number of remaining items} \notag \\
  &= F + A\big((n - m) + (m-G)\big)  , \notag \\
  &= F + An - AG , \notag
  \end{align}
where $n-m$ is the number of items not considered in the pass, and $m-G$ is the number of items not classified by the pass.
This is solved by $A = F/G$. Thus
  \[ A = \frac{F}{G} = \frac{1 +(1 - q^m)\log_2 m}{mq^m + \frac{1}{p} \big(1 + mq^{m+1} - (m+1)q^m\big)} . \]
When optimised over $m$ a power of $2$, this achieves rates $H(p)/A$ of over $0.95$ for all $p \leq 1/2$.
  
As before, setting $m =1$ recovers individual testing, and we indeed get $A=1$.
Setting $m = 2$, we have
  \[ A = \frac{1 +(1 - q^2)}{\frac{1}{p} \big(1 + 2q^{3} - 3q^2\big) + 2q^2} = \frac{2-q^2}{1+q} = \frac{1 + 2p - p^2}{2-p} . \]
We have $A < 1$, therefore outperforming individual testing, when $p \leq p^* = (3 - \sqrt{5})/2$, recovering a result of \cite{ungar}.

\subsection{General algorithm: average-case analysis} \label{sec:gen}

When considering analysis in the average case, the rate for some $p < 1/4$ can be improved by considering $m$ to be any integer, not just a power of $2$ (see the right-hand side of Fig.~\ref{mainfig}). We now explain how to perform binary splitting in this general case. We write $2^a = \lfloor m \rfloor_2$ and $b = m - 2^a$, so that $m = 2^a + b$ for integers $a$ and $b$ with $0 \leq b < 2^a$.

\begin{algorithm} \label{gen}
We wish to binary split a set $\mathcal B$ of size $m$ that contains at least one defective. We use a Huffman tree for the uniform distribution $(\frac1m,\frac1m,\dots,\frac1m)$. The $k$th test pool consists of the remaining items that have $k$th bit of their Huffman codeword equal to $0$; if the test is positive, the untested items are removed, while if the test is negative, the tested items are removed. When one item remains, it is defective.
\end{algorithm}

\smallskip

It is a standard result that Huffman coding for the uniform distribution results in $2^a - b = m - 2b$ items with wordlength $a$ and the remaining $2b$ items with wordlength $a + 1$. It will be convenient for the purposes of a later proof for the items of $\mathcal B$ in label order to be given Huffman codewords that are in lexicographic order, and that the shorter words are given to the first $m - 2b$ of the items. This means that we discover the status of the first defective item in $\mathcal B$ and all the preceding nondefective items.

It can be verified without too much difficulty that using $m$ that is not a power of $2$ does not improve the performance of Algorithm \ref{mainalg} in the zero-error case, but for reasons of space we do not  give the calculations here.

It appears that Algorithm \ref{mainalg} when used with Algorithm \ref{gen} for binary splitting is equivalent to an algorithm studied by Zaman and Pippenger \cite{zaman}. Their algorithm was defined in terms of optimal prefix-free codes for the geometric and truncated geometric distributions, and they used known results on such codes to prove that this algorithm is optimal among a set of algorithms called `nested algorithms'. They also studied the asymptotics of the quantity $\lim_{p \to 0} \lim_{n\to \infty} \bar T/k$, which, in our notation, corresponds to $\lim_{p\to 0} A/p$ where $A$ is the average-case achievable aspect ratio. They did not look at the rate for all $p$ or consider small-error testing.

We now analyse the average-case number of tests $\bar T$ of Algorithm \ref{mainalg} for arbitrary $m$.
Again, we look at the outcomes for a pass through step 2 of Algorithm \ref{mainalg}.
\begin{enumerate}
  \item[a)] With probability $q^m$, all items in the test are negative, and we discover their nondefective statuses with $1$ test.
  \item[b1)] With probability $1 - q^{m-2b}$ there is at least one defective in the first $m - 2b$ items in the test. Let $j$ be the first-numbered defective in the set. We discover defective status of item $j$ and the nondefective statuses of items $1,2,\dots,j-1$ in $a+1$ tests.
  \item[b2)] With probability $q^{m-2b} - q^m$ there are no defectives in the first $m - 2b$ items in the test, but there is at least one defective in the test. Let $j$ be the first-numbered defective in the set. We discover defective status of item $j$ and the nondefective statuses of items $1,2,\dots,j-1$ in $a+2$ tests.
\end{enumerate}

The expected number of tests in one pass through step 2 is
  \[ F = q^m\cdot 1 + (1 - q^{m-2b})(a+1)  + (q^{m-2b} - q^m)(a+2) . \]
The expected number of items whose status we discover is the same as before,
  \[ G = mq^m + \frac{1}{p} \big(1 + mq^{m+1} - (m+1)q^m\big)  . \]
The same argument as before shows that $A = F/G$, and it's easy to check, as noted in \cite{zaman}, that
  \begin{equation} \label{mstar}
  m = \left\lceil -\frac{\log(1+q)}{\log q} \right\rceil = \left\lceil -\frac{\log(2-p)}{\log(1-p)} \right\rceil
  \end{equation}
is the optimal value of $m$. The average number of tests required is $\bar T = An$.

\subsection{Small-error rate}

We now wish to prove Theorem \ref{mainthm2} by converting the above average-case result into a small-error result. To do this we will use a concentration of measure argument.

\begin{IEEEproof}[Proof of Theorem \ref{mainthm2}]
Let $\bar T$ be the average number of tests used, as calculated in the previous section. We will show that there is concentration of measure of the actual number of tests required, which, for any $\delta > 0$ is in the interval $T \in$ $\big((1-\delta)\bar T, (1+\delta)\bar T\big)$ with probability tending to $1$ as $n \to \infty$.

We then define an algorithm using $(1 + \delta)\bar T$ tests as follows. We run Algorithm \ref{mainalg} with the optimal value of $m$ as in \eqref{mstar}. If the algorithm takes fewer than $(1 + \delta)\bar T$ tests, we add extra arbitrary tests until it does, while if it take more than $(1 + \delta)\bar T$ tests, we stop at that point and guess the defective set arbitrarily. Clearly we can only make an error in the second case, and, once we have proved concentration of measure, that probability can be made arbitrarily small. By picking $\delta > 0$ sufficiently small, we ensure that all aspect ratios up to $A$ as in Theorem \ref{mainthm2} are achievable.

To prove concentration, we use McDiarmid's inequality \cite{mcdiarmid}, which gives concentration of measure when a bounded difference property holds. Let
  $T(x_1,x_2,\dots,x_n)$
be the number of tests used by Algorithm \ref{mainalg} when $x_i =1$ denotes that item $i$ is defective and $x_i = 0$ denotes it is nondefective. The random variable counting the number of tests used is 
  $T = T(X_1,X_2, \dots, X_n)$, where the $X_i$ are independent Bernoulli$(p)$ random variables.
  
To see that we have the necessary bounded difference property, we claim that, for $x_1, x_2, \dots, x_i, x_i', \dots, x_n \in \{0,1\}$, we have
\[ \big| T(x_1,x_2,\dots,x_i,\dots,x_n) - T(x_1,x_2,\dots,x_i',\dots,x_n) \big| \leq 2m . \]
Note from Algorithms \ref{pow2} and \ref{gen} that we discover the status of items is in increasing order of their labels. Thus changing $x_i$ to $x_i'$ will only change the number of tests between the last defective before $i$ and the first defective after $i$; outside that interval, the algorithm proceeds exactly the same. Thus changing $x_i$ might effect the number of tests for the first set $\mathcal B$ that covers $i$ after the previous defective being discovered -- potentially an increase or decrease of $a + 1$ tests, which we can bound by $a + 1 \leq m$. The same thing could happen when reaching the next defective after $i$, for a potential decrease of $a+1 \leq m$ tests again. This proves the bounded difference claim. McDiarmid's inequality then says that
\[ \mathbb P( |T - \bar T| > \delta \bar T ) \leq \exp \left( - \frac{2(\delta\bar T)^2}{n(2m)^2} \right) \leq\exp \left( - \frac{\delta^2H(p)^2}{2m^2} \,n \right) ,  \]
where we used the fact that $\bar T \geq H(p)n$.

Thus we have the desired concentration, and we are done.
\end{IEEEproof}


\section{Acknowledgements}

The author thanks Oliver Johnson and Jonathan Scarlett for useful comments.



\bibliographystyle{IEEEtran}
\bibliography{bibliography}

\end{document}